\documentclass[letterpaper,10pt]{article}
\usepackage{osameet3}
\usepackage[amssymb]{SIunits}
\usepackage[cmex10]{amsmath} 
\usepackage[latin1]{inputenc}
\usepackage{xcolor}
\usepackage{amsmath, amssymb,amsthm}
\usepackage{bm}
\usepackage{lipsum}
\usepackage{enumitem}
\usepackage[colorlinks=true,bookmarks=false,citecolor=blue,urlcolor=blue]{hyperref} 

\usepackage{graphicx}
\usepackage[caption=false, font=footnotesize]{subfig}
\usepackage{xspace} 
\usepackage[font=footnotesize]{caption}

\usepackage{multirow}
\usepackage{multicol}
\usepackage{floatrow}
\usepackage{wrapfig}
\newfloatcommand{capbtabbox}{table}[][\FBwidth]

\usepackage{titlesec}
\titlespacing\subsection{0pt}{4pt plus 2pt minus 2pt}{3pt plus 2pt minus 2pt}

\newcommand{\vect}[1]{\boldsymbol{#1}}

\newcommand{\imag}{\ensuremath{j}}
\renewcommand{\vect}[1]{\ensuremath{\mathbf{#1}}}
\newcommand{\mat}[1]{\ensuremath{\mathbf{#1}}}

\begin{document}

\title{Model-Based Machine Learning for Joint Digital Backpropagation
and PMD Compensation}

\newcommand{\aff}[1]{\textsuperscript{(#1)}}

\author{%
Christian H\"ager\aff{1}, %
Henry D.~Pfister\aff{2}, %
Rick M.~B\"utler\aff{3},\\ %
Gabriele Liga\aff{3}, %
Alex Alvarado\aff{3}%
}%
\address{\aff{1}{Department of Electrical Engineering, Chalmers
University, Gothenburg, Sweden}\\
\aff{2}{Department of Electrical and Computer Engineering, Duke
University, Durham, USA}\\
\aff{3}{Department of Electrical Engineering, Eindhoven University of
Technology, Eindhoven, The Netherlands}%
}%
\email{christian.haeger@chalmers.se}

\copyrightyear{2020}

\vspace{-0.2cm}


\begin{abstract}
	We propose a model-based machine-learning approach for
	polarization-multiplexed systems by parameterizing the split-step
	method for the Manakov-PMD equation. This approach performs
	hardware-friendly DBP and distributed PMD compensation with
	performance close to the PMD-free case.
\end{abstract}

\vspace*{-0.05cm}

\ocis{(060.0060) Fiber optics and optical communications, (060.2330)
Fiber optics communications.}

\vspace*{-0.25cm}

\section{Introduction}

The traditional application of machine learning to physical-layer
communication replaces individual signal processing blocks (e.g.,
equalization or decoding) by neural networks (NNs) with the aim of
learning better-performing algorithms (or less complex ones) through
data-driven optimization. More generally, one may regard the entire
communication system design as an end-to-end reconstruction task and
jointly optimize transmitter and receiver NNs \cite{OShea2017}. Both
traditional \cite{Shen2011, Jarajreh2015, Estaran2016} and end-to-end
learning \cite{Shen2018ecoc, Jones2018, Karanov2018} have received
considerable attention for optical fiber systems. However, the
reliance on NNs as universal (but sometimes poorly understood)
function approximators makes it difficult to incorporate existing
domain knowledge or interpret the obtained solutions. 

Rather than relying on NNs, a different approach is to start from an
existing model and parameterize it. For fiber-optic systems, this can
be done for example by considering the split-step method (SSM) for
numerically solving the nonlinear Schr\"odinger equation (NLSE).  By
viewing all chromatic-dispersion steps as general linear functions,
one obtains a parameterized model similar to a multi-layer NN
\cite{Haeger2018ofc}. Compared to standard ``black-box'' models, this
approach has several advantages: it leads to clear hyperparameter
choices (such as the number of layers/steps); it provides good
initializations for a gradient-based optimization; and it allows one
to inspect the learned solutions in order to understand \emph{why}
they work well, thereby providing significant insight into the problem
\cite{Haeger2018ofc, Haeger2018isit, Lian2018itw}. 

In this paper, we extend the model-based approach proposed in
\cite{Haeger2018ofc} to polarization-multiplexed (PM) systems by
parameterizing the SSM for the Manakov-PMD equation. This leads to a
multi-layer model alternating complex-valued $2 \times 2$
multiple-input multiple-output finite impulse response (MIMO-FIR)
filters with nonlinear Kerr operators. The complexity of the filters
can be reduced by decomposing them into separate filters for each
polarization followed by memoryless rotation matrices
\cite{Haeger2019ecoc}. As an application, we consider joint digital
backpropagation (DBP) and polarization-mode dispersion (PMD)
compensation, similar to \cite{Goroshko2016, Czegledi2016,
Czegledi2017, Liga2018}. Compared to previous work, the employed model
uses hardware-friendly time-domain implementations
\cite{Fougstedt2017,Fougstedt2018ecoc} and does not assume any
knowledge about the particular PMD realizations along the link (i.e.,
differential group delays (DGDs) and polarization states), nor any
knowledge about the total accumulated PMD. We show that our model
converges reliably to a solution with performance close to the case
where PMD is absent from the link. 

\section{Supervised Machine Learning}

We start by reviewing the standard supervised learning setting for
feed-forward NNs. A feed-forward NN with $M$ layers defines a mapping
$\vect{y} = f_\theta(\vect{x})$ where the input vector $\vect{x} \in
\mathcal{X}$ is mapped to the output vector $\vect{y} \in \mathcal{Y}$
by alternating between affine transformations $\vect{z}^{(i)} =
\vect{W}^{(i)} \vect{x}^{(i-1)} + \vect{b}^{(i)}$ and pointwise
nonlinearities $\vect{x}^{(i)} = \phi(\vect{z}^{(i)})$ with
$\vect{x}^{(0)} = \vect{x}$ and $\vect{x}^{(M)} = \vect{y}$. The
parameter vector $\theta$ comprises all elements of the weight
matrices $\vect{W}^{(1)}, \dots, \vect{W}^{(M)}$ and vectors
$\vect{b}^{(1)},\dots,\vect{b}^{(M)}$. Given a training set $S\subset
\mathcal{X} \times \mathcal{Y}$ that contains a list of input--output
pairs, training proceeds by minimizing the empirical loss
$\mathcal{L}_S (\theta) \triangleq \frac{1}{|S|}
\sum_{(\vect{x},\tilde{\vect{y}})\in S} \ell \big(
f_\theta(\vect{x}),\tilde{\vect{y}})$, where
$\ell(\vect{y},\tilde{\vect{y}})$ is the per-sample loss associated
with returning the output $\vect{y} = f_\theta(\vect{x})$ when
$\tilde{\vect{y}}$ is correct. When the training set is large, one
typically optimizes $\theta$ using a variant of stochastic gradient
descent (SGD). In particular, mini-batch SGD uses the parameter update
$\theta_{t+1} = \theta_t - \alpha \nabla_\theta \mathcal{L}_{B_t}
(\theta_t)$, where $\alpha$ is the step size and $B_t \subseteq S$ is
the mini-batch used in the $t$-th step. 

Supervised machine learning is not restricted to NNs and learning
algorithms such as SGD can be applied to other function classes as
well. In fact, prior to the current revolution in machine learning,
communication engineers were quite aware that system parameters (such
as filter coefficients) could be learned using SGD. It was not at all
clear, however, that more complicated parts of the system architecture
could be learned as well. For example, in practice, PMD can be
compensated by choosing the function $f_\theta$ as the convolution of
the received signal with the impulse response of a complex-valued $2
\times 2$ MIMO-FIR filter, where $\theta$ are the filter coefficients.
For a particular choice of the loss function $\ell$, applying SGD then
maps into the well-known constant modulus algorithm \cite{Savory2008}.

\section{Model-Based Machine Learning for Polarization-Multiplexed
Systems}

\newcommand{\pTX}{\mathcal{T}_{\theta}}
\newcommand{\pRX}{\mathcal{R}_{\theta}}
\newcommand{\pCH}{\mathcal{C}_{\theta}}

\begin{figure}[t]
	\centering
		\includegraphics{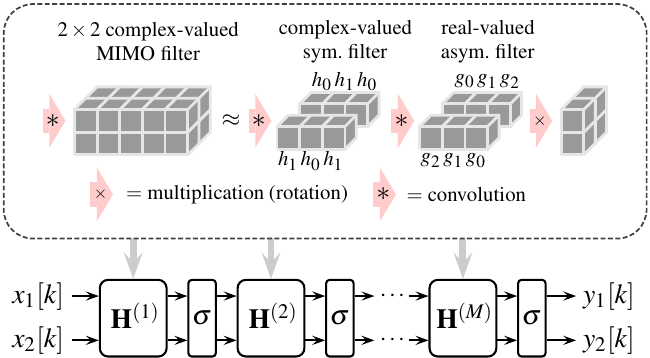}
		$\quad$
		\includegraphics{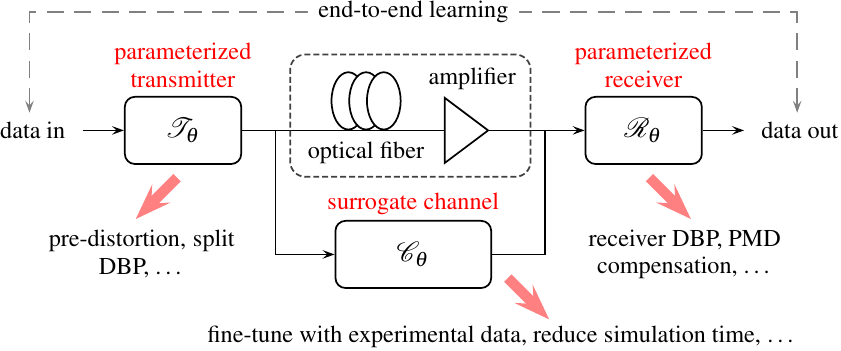}
		\caption{Left: Block diagram of the proposed model $\vect{y} =
		f_{\theta}(\vect{x})$ based on
	the SSM for the Manakov-PMD equation, where all
	filters are free parameters $\theta =
	\{\mat{H}^{(1)}, \dots, \mat{H}^{(M)}\}$ and $\sigma(\vect{x}) = \vect{x} e^{\imag \gamma \frac{8}{9} \delta
\|\vect{x}\|^2}$. The framed box illustrates the 
	MIMO filter decomposition used for the numerical results. Right:
	Potential applications in the context of end-to-end
	learning \cite{OShea2017}, where $\pTX, \pRX, \pCH$ are
	parameterized functions.}
	\label{fig:model_based}
\end{figure}

The evolution of PM signals over single-mode fiber can be described by
a set of coupled NLSEs that takes into account the interactions
between the two (degenerate) polarization modes. In birefringent
fibers where polarization states change rapidly along the link, an
appropriate approximation is given by the Manakov-PMD equation
\cite{Wai1991} $\frac{\partial}{\partial z} \vect{u} = \hat{\vect{D}}
\, \vect{u} + \imag \gamma \frac{8}{9} \| \vect{u} \|^2 \, \vect{u}$,
where $\vect{u} = \vect{u}(t,z) = (u_1(t,z), u_2(t,z))^\top$ is the
Jones vector comprising the complex baseband signals in both
polarizations, $\gamma$ is the nonlinear Kerr parameter, and
$\hat{\vect{D}}$ is a linear operator that models attenuation,
chromatic dispersion (CD), and PMD, see, e.g.,
\cite[Eq.~(1)]{Ip2010a}. 

Similar to the NLSE, the Manakov-PMD equation has no general
closed-form solutions but can be solved numerically using the SSM. The
SSM alternates linear and nonlinear steps, where the nonlinear steps
are described by $\sigma(\vect{x}) = \vect{x} e^{\imag \gamma
\frac{8}{9} \delta \|\vect{x}\|^2}$ for $\vect{x} \in \mathbb{C}^2$
and step size $\delta$. The linear steps can be approximated by $2
\times 2$ MIMO-FIR filters in discrete time, which leads to the model
shown in Fig.~\ref{fig:model_based} (left). The proposed approach is
to fully parameterize this model by regarding all filter coefficients
in all steps as free parameters. In the context of end-to-end
learning, as illustrated in Fig.~\ref{fig:model_based} (right), such a
model can serve for example as the basis for a fine-tuned channel
model based on experimental data, using techniques similar to
\cite{OShea2019}. In this paper, we focus on receiver-side DBP, where
the model can learn and apply arbitrary filter shapes to PM signals in
a distributed fashion. The model can also be used at the transmitter
to learn optimized pre-distortions \cite{Essiambre2005, Roberts2006}
or split-DBP \cite{Lavery2016a}. 

For multi-layer models, it is important to simplify the individual
steps as much as possible to limit the overall complexity
\cite{Haeger2019ecoc}. This is especially important since PMD is a
time-varying impairment that requires adaptive filtering in practice.
We therefore decompose each MIMO filter into three components as shown
in Fig.~\ref{fig:model_based} (left): (1) a complex-valued symmetric
filter accounting for CD, (2) a real-valued asymmetric filter with
``flipped'' coefficients in different polarizations to approximate
fractional-delay filters accounting for DGD, (3) a memoryless $2
\times 2$ rotation matrix. Such a decomposition considerably reduces
the number of parameters and required hardware operations compared to
a $2 \times 2$ MIMO filter. For the rotations, we assume that each
matrix can be represented as $\left(\begin{smallmatrix} a & -b^* \\ b
	& a^*
\end{smallmatrix}\right)$, where $a, b \in \mathbb{C}$, i.e., $4$ real
parameters per step. While this does not impose a unitary constraint
on the rotation, we found that it leads to better SGD training
compared to other parameterizations used in, e.g., \cite{Liga2018}.

\section{Numerical Results}

We consider the simulation setup in \cite{Czegledi2016} with slightly
adjusted parameters. In particular, we assume single-channel
transmission of a $32$ Gbaud PM signal (Gaussian symbols, root-raised
cosine, roll-off $0.01$) over $10 \times 100$ km of fiber ($\alpha =
0.2$ dB/km, $\beta_2 = -21.683$ ps$^2$/km, $\gamma = 1.2$ rad/W/km),
where optical amplifiers (noise figure $4.5$ dB) compensate for
attenuation after each span. Forward propagation is simulated with
$1000$ uniform steps per span (StPS) and $192$ GHz simulation
bandwidth. PMD was emulated at every step with a frequency-dependent
Jones matrix $\mat{R}^{(i)} \mat{J}^{(i)}(\omega)$, $i = 1,\dots, N =
10000$, where $\mat{J}^{(i)}(\omega) = \text{diag}(e^{-\imag \omega
\frac{\tau_i}{2}}, e^{\imag \omega \frac{\tau_i}{2}} )$ is a
first-order PMD matrix with DGD $\tau_i$ and $\mat{R}^{(i)}$ is a
unitary rotation matrix. For generating a particular PMD realization
$\{\tau_i, \mat{R}^{(i)}\}_{i=1}^{N}$ for the entire link, we use the
same approach as described in \cite{Czegledi2016} with PMD parameter
$0.2\,$ ps/$\sqrt{\text{km}}$. At the receiver, the signal was
low-pass filtered ($64$ GHz bandwidth) and downsampled to $2$
samples/symbol for further processing. We started by training a
$4$-StPS model without rotations or DGD filters, assuming no PMD
(i.e., $\tau_i =0$, $\mat{R}^{(i)} = \left(
\begin{smallmatrix} 1 & 0\\0 & 1
\end{smallmatrix} \right)$ for all $i$). 
The model has $41$ steps based on the symmetric SSM, where
the last step is a linear ``half-step'' without nonlinearity, followed
by a matched filter (MF) and phase-offset correction. Mean-squared
error loss is employed,

\begin{wrapfigure}[18]{r}{8.5cm}
	\vspace{-0.25cm}
	\centering
	\includegraphics{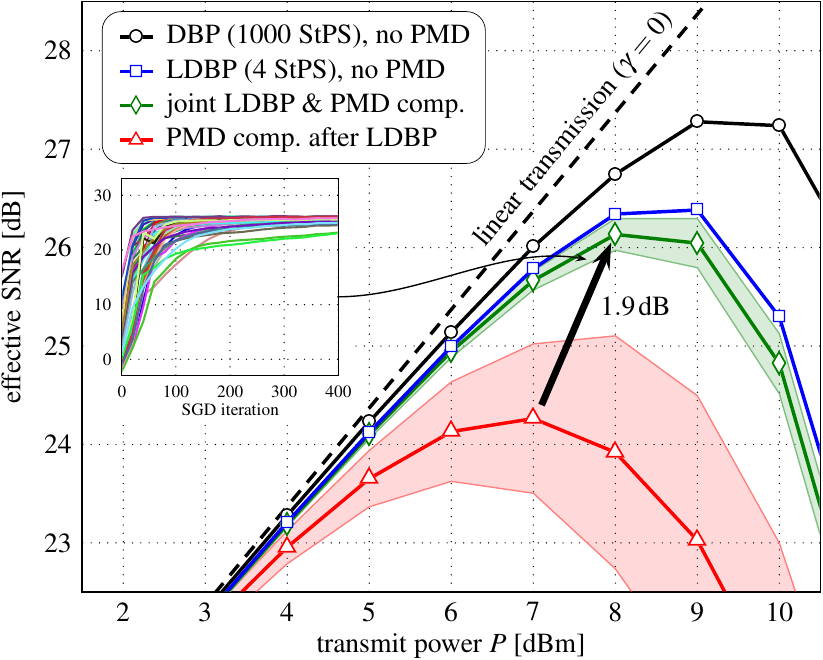}
	\vspace*{-0.20cm}
	\caption{Results for $40$ PMD realizations (shaded areas are
	std.~deviations).}
	\label{fig:effective_snr}
\end{wrapfigure}

\noindent which is essentially the same as standard learned DBP (LDBP)
\cite{Haeger2018ofc}. The solution is then used for two receiver
setups accounting for PMD, where the corresponding performance serves
as an upper bound, see Fig.~\ref{fig:effective_snr} (blue squares;
``ideal'' DBP with $1000$ StPS is also shown by the black circles).
The first setup uses the same LDBP followed by the inverse of the PMD
transfer function matrix $\mat{J}(\omega) = \prod_{i=1}^N
\mat{R}^{(i)}\mat{J}^{(i)}(\omega)$, prior to the MF. No additional
training is performed.  Fig.~\ref{fig:effective_snr} shows the mean
effective SNR averaged over $40$ PMD realizations (red triangles,
shaded areas indicate standard deviation). As expected, PMD reduces
the effectiveness of LDBP, which is similar to standard DBP.  Next,
the model was extended with rotations and $5$-tap DGD filters.  The CD
filters are frozen to the LDBP solution during training, giving only
$9$ free parameters per step. The rotations are randomly initialized
and DGD filters are initialized to $(0,0,1,0,0)$. The model is trained
for $1500$ iterations with the Adam optimizer \cite{Kingma2014},
learning rate $0.0005$, and batch size $50$. The results in
Fig.~\ref{fig:effective_snr} (green diamonds) show that the mean
performance for the same $40$ PMD realizations is improved by $1.9$
dB, which is only $0.2$ dB away from LDBP without PMD. Moreover, the
standard deviation is reduced by more than a factor of $6$. The
training curves in Fig.~\ref{fig:effective_snr} demonstrate a quick
convergence behavior for all considered PMD realizations, where the
starting SNR for the optimization is random due to the initialization
of the rotations.

\section{Conclusions}

We have proposed a multi-layer machine-learning model for PM systems
based on parameterizing the SSM for the Manakov-PMD equation. This
model can be used for end-to-end learning as part of a parameterized
transmitter or receiver implementation (or both), or as the basis for
a fine-tuned channel model. In this paper, we have considered
receiver-side DBP and PMD compensation, demonstrating performance
close to the PMD-free case, without assuming any knowledge about the
PMD realizations along the link or the total accumulated PMD. 

\vspace{0.2cm}
{\scriptsize 
\noindent{\bf{Acknowledgements}}:
This work is part of a project that has received funding from the
European Union's Horizon 2020 research and innovation programme under
the Marie Sk\l{}odowska-Curie grant agreement No.~749798. The work of
H.~D.~Pfister was supported in part by the National Science Foundation
(NSF) under Grant No.~1609327.  The work of A.~Alvarado and G.~Liga
has received funding from the European Research Council (ERC) under
the European Union's Horizon 2020 research and innovation programme
(grant agreement No.~757791). Any opinions, findings, recommendations,
and conclusions expressed in this material are those of the authors
and do not necessarily reflect the views of these sponsors.

}

\vspace{-0.3cm}

\appendix

\newif\iffullbib
\fullbibfalse

\newcommand{\jlt}{J.~Lightw.~Technol.}
\newcommand{\ope}{Opt.~Exp.}
\newcommand{\tit}{IEEE Trans.~Inf.~Theory}
\newcommand{\tc}{IEEE Trans.~Comm.}
\newcommand{\ofc}{Proc.~OFC}
\newcommand{\ecoc}{Proc.~ECOC}
\newcommand{\ita}{Proc.~ITA}
\newcommand{\scc}{Proc.~SCC}


\iffullbib

\else

\fi

\end{document}